\documentclass[aps,pra,showpacs,twocolumn]{revtex4}
\usepackage{amsmath, amssymb}
\usepackage{graphicx}
\usepackage{subfigure}

\begin{document}

\title{Photon emission by an atom in a lossy cavity}

\author{C. Di Fidio}

\author{W. Vogel}
\affiliation{Arbeitsgruppe Quantenoptik, Institut f\"ur Physik,
Universit\"at Rostock, D-18051 Rostock, Germany}

\author{M. Khanbekyan}

\author{D.-G. Welsch}
\affiliation{Theoretisch-Physikalisches Institut,
Friedrich-Schiller-Universit\"at Jena, Max-Wien-Platz 1,
D-07743 Jena, Germany}

\date{December 11, 2007}

\begin{abstract}

The dynamics of an initially excited two-level atom in a lossy cavity is studied by using the quantum trajectory method.
Unwanted losses are included, such as photon absorption
and scattering by the cavity mirrors and spontaneous emission of the atom. 
Based on the obtained analytical solutions, it is shown that the shape of the extracted 
spatiotemporal radiation mode sensitively depends on the atom-field interaction. 
In the case of a short-term atom-field
interaction we show how different pulse
shapes for the field extracted from the cavity can be controlled by the interaction time.

\end{abstract} 


\maketitle

\section{Introduction}
\label{introduction}

A single atom interacting with a
quantized radiation-field mode 
in a high-$Q$ optical cavity plays 
an important role in quantum optics
not only due to its conceptual relevance,
but also because it appears as a
basic element in various
schemes, such as in the field of
quantum information science (for a review see, e.g.,
Refs.~\cite{HarocheRaimond,Walther1, Walther2, Nielsen}).
Cavity quantum electrodynamics (QED)
has been used for
the generation and processing of nonclassical
radiation, as, for example, in the
single-atom maser~\cite{Haroche:347,Gallas:414,Meschede:551, Rempe:353,Brune:1899} or in the optical domain~\cite{Thompson:1132}.
The quantum control of single-photon 
emission from an atom in a cavity
for generating single-photon Fock states 
on demand
has been realized~\cite{Monroe:238}, 
and single-photon Fock state generation
of high efficiency has been
a key requirement in various applications such as
quantum cryptography~\cite{Bennett:2724, Luetkenhaus:52304}
or quantum networking
for distribution and processing of quantum
information~\cite{Cirac:3221, Knill:46}.
Recently, single-photon sources operating
on the basis of adiabatic passage with just one atom
trapped in a high-$Q$ optical cavity
have been realized~\cite{Parkins:3095,Hennrich:4872, Mckeever:1992, Hijlkema:253}.
In this way, the generation of single photons of
known circular polarization has been possible \cite{Wilk:063601}.
Moreover, the adjustment of the spatiotemporal profile of
single-photon pulses has been achieved
\cite{Kuhn:067901, Keller:1075}.

In view of the widespread applications of
cavity-assisted single-photon sources, it is of great
importance to carefully study the quantum state of
the field escaping from a cavity.
Let us consider the simplest case of a two-level atom that
near-resonantly interacts with a narrow-band
cavity-field mode. On a time scale that is
sufficiently short compared to the inverse
bandwidth of the mode, the radiative and non-radiative
cavity losses may be disregarded, and the
atom-field dynamics can be described by the
familiar Jaynes-Cummings model~\cite{Jaynes:89}.
Clearly, for longer times, the atom-cavity system can no longer
be regarded as being a closed system, and the losses must
be taken into account. Since the wanted outgoing field
represents, from the point of view of the atom--cavity system,
radiative losses, the study of the input--output problem
necessarily requires inclusion in the theory of the effects
of losses.
Such a system, consisting of a two-level atom 
interacting with a single mode of a lossy
cavity, has been widely considered
in the past decades. 
Some of the initial theoretical works
treating the effects of losses 
on the Jaynes-Cummings dynamics
can be found in Refs.~\cite{Barnett:2444,Filipowicz:3077,Puri:3433,Kuklinski:3175,Risken:346,Cirac:4541,Quang:6092,Alsing:13,Banacloche:2221}. For a review on this topic
see, e.g., Refs.~\cite{Shore:1195,Barnett:2033}.
Anyway, a detailed characterization of the
cavity output field in such a system still presents
some open questions of significant interest, as, for example,
the control of the pulse shape of the emitted photon. 

There are primarily two approaches to this
problem, which are based on either quantum field theory
or quantum noise theory.
In quantum field theory, the system is
commonly described on the basis of Maxwell's equations
as used in macroscopic QED~\cite{Knoell:1,VogelWelsch}.
It has been shown that an approximate description of the fields inside and outside a cavity can be formulated in terms
of quantum Langevin equations and input-output relations~\cite{Knoell:543, Plank:1791}.
Macroscopic QED can also be used to
study effects of
unwanted losses, such as scattering and absorption losses caused by the cavity mirrors~\cite{Khanbekyan:053813,Semenov:033803,Semenov:013807}.
More recently, the photon emission by an excited atom in a cavity has been analyzed by the method of macroscopic QED~\cite{Khanbekyan:quant-ph}.
By using a source-quantity representation
of the electromagnetic field, the properties of the outgoing field are investigated. In such an approach the field inside and outside the cavity is combined in a unique radiation mode, without regarding the fields inside
and outside the cavity as representing independent
degrees of freedom.

Conversely, in quantum noise theory the fields inside and
outside a cavity are regarded as representing independent
degrees of freedom~\cite{Collett:1386, Gardiner:3761, GardinerZoller}.
Accordingly, such a theory is based on discrete and continuous
mode expansions of the fields inside and outside the
cavity, respectively. Thus the operators of the intracavity and external fields are regarded as commuting quantities.
The continuum of the external modes is regarded as playing
the role of a dissipative system. Its effect on
the dynamics of the intracavity modes can be treated by
quantum Langevin equations, or, alternatively, by master
equations~\cite{Haake, Louisell, Davies}.
For obtaining their solution one can apply,
for example, the quantum
trajectory method~\cite{Dalibard:580, Dum:4382,Carmichael}. 

In the present paper we consider a two-level atom
interacting with a lossy cavity, within the framework
of quantum noise theory,
giving particular emphasis 
to the derivation of the pulse shape of the emitted photon.
The dynamical evolution of the open quantum system under study is described
by a master equation for the reduced density operator of the
atom-cavity system. The effects of unwanted losses, such as spontaneous emissions of the two-level atom out the side of the cavity
and photon absorption and scattering by the cavity mirrors, are also
taken into account.
For an initially excited two-level atom in an empty cavity,
we solve the master equation analytically by using the quantum trajectory method.
In order to characterize the cavity output field,
described by a single-photon spatiotemporal mode,
we connect the probability to measure a photon
in this mode with the photodetection probability as given
by the quantum trajectory theory.
This allows us to 
derive the shape of the mode of the extracted
cavity field, which shows a clear 
mapping of the intracavity-field dynamics onto the mode of the output field. 
Moreover, the probability of the outgoing mode to carry
a single-photon Fock state is calculated.
After considering the case of a continuing atom-field
interaction, we also analyze the short-term atom-field interaction. It is shown that, by changing the interaction time
of the atom with the cavity, the mode structure of the
outgoing field results in different pulse shapes of the single-photon wave packet. This opens 
possibilities to control the shape of the pulse.

The paper is organized as follows.
In Sec.~\ref{section:1} the master equation 
describing the dynamics of the atom-cavity
system is introduced,
and the problem is solved
analytically by using the quantum trajectory method.
In Sec.~\ref{section:2}, giving a description of the single-photon wave packet in terms of spatiotemporal
mode functions, the shape of the mode of the extracted
cavity field is obtained.
In Sec.~\ref{section:3} we analyze short-term atom-field interactions and obtain different shapes of the mode of the 
extracted field.
A summary and some concluding remarks are given in Sec.~\ref{conclusions}.

\section{Damped atom-field dynamics}
\label{section:1}

\begin{figure}
\includegraphics[width=7cm]{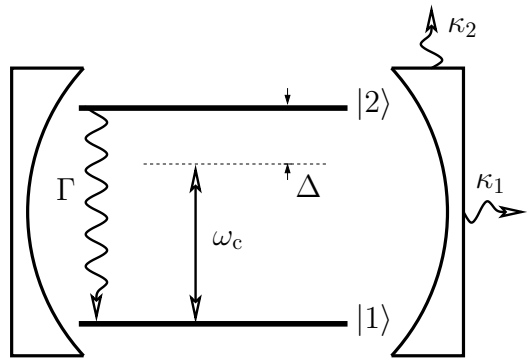}
\caption{The cavity mode of frequency $\omega_{\rm c}$ is detuned by $\Delta$ from the two-level atomic transition
frequency $\omega_{\rm a} = \omega_{\rm c} + \Delta$. 
$\kappa_1$ and $\kappa_2$ are
the photon escape rate of the cavity and the
cavity mirrors' absorption and scattering rate, respectively.
$\Gamma$ is the dipole relaxation rate from level $|2\rangle$ to $|1\rangle$.}
\label{fig:figure_pra_1}
\end{figure}

In this section we analyze the dynamics of 
the system under scrutiny 
starting from a master equation
and solving it by using the quantum trajectory theory.
We consider a single two-level atomic transition
of frequency  $\omega_a$ coupled to a cavity mode 
of frequency $\omega_c$.
The cavity mode is detuned by $\Delta$ from the 
two-level atomic transition frequency, $\omega_{\rm a} = \omega_{\rm c} + \Delta$, and 
is damped by losses through the partially transmitting cavity mirrors, cf. Fig.~\ref{fig:figure_pra_1}. 
In addition to the wanted outcoupling of the field,
the photon can be spontaneously emitted out the side
of the cavity into modes other than the one
which is preferentially coupled to the resonator.
Moreover, the photon may be absorbed or scattered
by the cavity mirrors.

Treating the dissipation due to the cavity losses
in a standard way~\cite{Haake, Louisell, Davies}, the
dynamical evolution of the
reduced density operator $\hat \rho(t)$ of the atom and 
the cavity field
is described by the following master equation 
\begin{eqnarray}
\frac{d\hat\rho(t)}{dt} \!\!\!&=&\!\!\! \frac{1}{i\hbar} \! \left[\hat H,\hat\rho(t)\right] \!+\! \sum_{i = 1}^2
 \frac{\kappa_i}{2} \!
\left[ 2\hat a \hat\rho(t) \hat a^\dagger \!-\!
  \hat a^\dagger \hat a \hat\rho(t) \!-\! \hat\rho(t) \hat a^\dagger \hat a \right] \nonumber \\
&+&\frac{\Gamma}{2} \left[ 2\hat A_{12} \hat\rho(t) \hat A_{21} -
  \hat A_{22} \hat\rho(t)  - \hat\rho(t)  \hat A_{22} \right] .
\label{eq:master_1}
\end{eqnarray}
Here $\kappa_1$ and $\kappa_2$ are
the photon escape rate of the cavity and the
cavity mirrors' absorption and scattering rate, respectively.
We denote by $\Gamma$ the spontaneous 
emission
rate of the two-level atom.
The Hamiltonian that describes the atom-cavity 
interaction
is given, in the
rotating-wave approximation, by
\begin{equation}
\hat H = \hbar{g} \left( \hat{a} \hat A_{21} 
+ \hat a^\dagger \hat A_{12}\right) + \hbar \Delta {\hat A}_{22}
   \, ,
\label{eq:JC_hamiltonian}
\end{equation}
where $\hat a$ and $\hat a^\dagger$ are annihilation and creation operators
for the cavity field, respectively, and 
$\hat A_{ij} = |i\rangle \langle j|$ ($i,j = 1, 2$),
where $|1\rangle$ and $|2\rangle$ are the 
two atomic energy eigenstates.
Moreover, $g$ is the atom-cavity coupling constant.
Here we are considering an
interaction picture with respect to $\hat H_0 - \hbar \Delta {\hat A}_{22}$, where ${\hat H}_0 = \hbar \omega_{\rm c} \hat a^\dagger \hat a + (1/2) \hbar \omega_{\rm a} ({\hat A}_{22} - {\hat A}_{11} )$.

For notational convenience, in the following we will identify the state $|a\rangle$ with the state
$|2,0\rangle$, the atom in the upper level, and no photon in the cavity.  The
state $|a\rangle$ will be considered to be the initial state of the system.  Moreover, we will indicate with $|b\rangle$ the state
$|1,1\rangle$, the atom in the lower level, and one photon in the cavity.  
Due to photon extraction through the cavity mirror, 
photon absorptions, or spontaneous emissions,
the quantum state of the atom-cavity system
is projected into the state $|c\rangle$, that indicates the
state $|1,0\rangle$, i.e., the atom in the lower level and no photon in the cavity.
It follows that the Hilbert space that describes the atom-cavity system under scrutiny is, in this model, 
simply spanned by the three vectors $|a\rangle$, $|b\rangle$
and $|c\rangle$.

To evaluate the time evolution of the system different
approaches can be used, see, for example, 
Refs.~\cite{Agarwal:1757,Puri:3610,Briegel:3311}.
Here we have found it quite convenient to use a quantum trajectory
approach~\cite{Dalibard:580, Dum:4382, Carmichael} to obtain 
our analytical solutions. 
In this approach the dynamical evolution
of the unnormalized state vector $| \bar{\psi} (t) \rangle$,
that describes the system at time $t$, 
is governed by a nonunitary Schr\"odinger equation with a
non-Hermitian Hamiltonian. The evolution generated by
this Schr\"odinger equation is
randomly interrupted, from time to time, by the action of
collapse, or jump, operators.

More precisely, in our specific case, considering the system prepared at time
$t_0 = 0$ in the state $|a\rangle$, to determine 
the state vector of the system at a later time $t$, 
assuming that no jump has occurred between time $t_0$ and $t$,  
we have to solve the nonunitary Schr\"{o}dinger
equation
\begin{equation}
i \hbar \frac{d}{dt} 
| \bar{\psi}_{\rm no} (t) \rangle  =\hat{H^{'}} \, | \bar{\psi}_{\rm no} (t) \rangle \, ,
\label{eq:schr_non_unitary} 
\end{equation}
where $\hat{H^{'}}$ is the non-Hermitian Hamiltonian
given by
\begin{equation}
\hat{H^{'}} = \hat H
- i \hbar \frac{\kappa}{2} \hat a^\dagger \hat a - i \hbar \frac{\Gamma}{2} \hat A_{22}\, ,
\label{eq:n_H_Hamiltonian}
\end{equation}
with $\hat H$ given by Eq.~({\ref{eq:JC_hamiltonian}),
and where we have defined
\begin{equation}
\kappa = \kappa_1 + \kappa_2 \, .
\label{eq:kappa}
\end{equation}
If no jump has occurred between time $t_0$ and $t$, the system evolves via Eq.~(\ref{eq:schr_non_unitary}) in the unnormalized state
\begin{equation}
| \bar{\psi}_{\rm no} (t) \rangle = \alpha(t) |a\rangle + \beta(t) |b \rangle \, .
\label{eq:psi_n}
\end{equation}
In this case the conditioned density operator for
the atom-cavity system is given by 
\begin{eqnarray}
\hat \rho_{\rm no} (t) &=& \frac{ | \bar{\psi}_{\rm no} (t) \rangle \langle
\bar{\psi}_{\rm no} (t) |}{\langle \bar{\psi}_{\rm no} (t) |\bar{\psi}_{\rm no} (t) \rangle}
\, .
\label{eq:rho_no}
\end{eqnarray}
Here we have used the word conditioned to stress the fact that 
this is the density operator at time $t$ one obtains conditioned to the fact that no jump has occurred between 
time $t_0$ and $t$.

The evolution governed by the nonunitary Schr\"odinger equation~(\ref{eq:schr_non_unitary})
is randomly interrupted by three kinds of jumps, 
$\hat J_1$, $\hat J_2$ and $\hat J_{\rm s}$
given by
\begin{eqnarray}
&& \hat J_i = \sqrt{\kappa_i} \hat a   ~~~(i = 1,2)\, , \label{eq:jump_i}  \\
&& \hat J_{\rm s} = \sqrt{\Gamma}\, \hat A_{12} \, .
\label{eq:jump_s}
\end{eqnarray}
The jump operators $\hat J_1$ and $\hat J_2$ are related 
to a photon extracted from the cavity and 
a photon absorbed or scattered by the mirrors, respectively.
The jump operator $\hat J_{\rm s}$ is related to 
a photon spontaneously emitted by the atom.
If a jump has occurred at time
$t_{\rm J}$, $t_{\rm J} \in (t_0, t]$, the
wavevector is found collapsed in the state $|c \rangle$ due to the action of one of the jump operators
\begin{eqnarray}
&& \hspace*{-0.5cm} \hat J_{i} \, | \bar{\psi}_{\rm no} (t_{\rm J}) \rangle \!=\! \sqrt{\kappa_i} \, \hat a
| \bar{\psi}_{\rm no} (t_{\rm J}) \rangle \rightarrow |c \rangle   ~~(i = 1,2), \label{eq:jump_op_i} \\
&& \hspace*{-0.5cm} 
\hat J_{\rm s} \, | \bar{\psi}_{\rm no} (t_{\rm J}) 
\rangle \!=\! \sqrt{\Gamma} \, \hat A_{12} | \bar{\psi}_{\rm no} (t_{\rm J}) \rangle \rightarrow |c \rangle \, . \label{eq:jump_op_s}
\end{eqnarray}

It is clear that in the problem under study we can have only one jump. Once the system collapses in the state $|c\rangle$ the nonunitary Schr\"odinger equation~(\ref{eq:schr_non_unitary}) simply keeps it there forever. In this case the conditioned density operator at time $t$ is given by
\begin{equation}
\hat \rho_{\rm yes} (t) = | c \rangle \langle c | \, ,
\label{eq:rho_yes}
\end{equation}
where we indicate with ``yes" the fact that a jump has occurred.

According to the quantum trajectory method, the density operator $\hat \rho (t)$ is obtained by performing an ensemble
average over the different conditioned density operators at
time $t$. In the present case, starting at time $t_0$ with the density operator $\hat \rho_0 = | a \rangle \langle a |$,
the ensemble average is performed over the two 
possible realizations (histories) ``yes" and ``no": 
\begin{equation}
\hat \rho (t) =  p_{\rm no}(t) \hat \rho_{\rm no} (t) +
  p_{\rm yes}(t) \hat \rho_{\rm yes} (t) \, .
\label{eq:rho_t}
\end{equation}
Here $p_{\rm no}(t)$ and $p_{\rm yes}(t)$ are the probability that between the initial time $t_0$ and time $t$ no jump and
one jump has occurred, respectively. Of course, 
$p_{\rm no}(t) + p_{\rm yes}(t) = 1$.
The density operator given by Eq.~(\ref{eq:rho_t})
tells us that the system at time $t$ is in a statistical mixture:  either no 
photon has escaped from the cavity or one (and only one) photon has escaped. 

To evaluate $p_{\rm no}(t)$ we use the method of the delay function~\cite{Dum:4382}. This method tells us that the probability $p_{\rm no}(t)$ is given by the square of the norm
of the unnormalized state vector:
\begin{equation}
p_{\rm no}(t) = \parallel | \bar{\psi}_{\rm no} (t) \rangle \! \parallel^2
= \langle \bar{\psi}_{\rm no} (t) |\bar{\psi}_{\rm no} (t) \rangle = 
|\alpha(t)|^2  + |\beta(t)|^2 \, .
\label{eq:p_no}
\end{equation}
From Eqs.~(\ref{eq:rho_t}) and (\ref{eq:p_no}) one obtains for the density operator $\hat \rho (t)$ the expression
\begin{eqnarray}
\hat \rho (t) &=& |\alpha(t)|^2 |a \rangle \langle a| + |\beta(t)|^2 |b \rangle \langle b| + \alpha(t)\beta^{*}(t) 
|a \rangle \langle b| \nonumber \\
&+& \alpha^{*}(t)\beta(t) |b \rangle \langle a| + |\gamma(t)|^2 |c \rangle \langle c| \, ,
\label{eq:rho_t_1}
\end{eqnarray}
where we have defined
\begin{equation}
|\gamma(t)|^2 \equiv p_{\rm yes}(t) = 
1 - \left[  |\alpha(t)|^2  + |\beta(t)|^2    \right] \, . 
\label{eq:gamma_def}
\end{equation}
The physical meaning of $|\alpha(t)|^2$, $|\beta(t)|^2$
and $|\gamma(t)|^2$ is clear. They represent the probability that at time $t$ the system can be found either in $|a\rangle$,
$|b\rangle$, or $|c\rangle$.
Moreover, from the master equation~(\ref{eq:master_1}),
together with Eq.~(\ref{eq:rho_t_1}),
one obtains
\begin{equation}
\frac{d |\gamma(t)|^2}{dt} = {\rm Tr} \left[ \frac{d \hat \rho (t)}{dt} |c\rangle 
	\langle c| \right] = \kappa |\beta(t)|^2 
+ \Gamma |\alpha (t)|^2 \, .
\label{eq:gamma_deriv}
\end{equation}
To better understand the meaning of Eq.~(\ref{eq:gamma_deriv}), it is useful to do the following consideration.
According to the quantum
trajectory theory, the probability for a jump, cf.
Eqs.~(\ref{eq:jump_i}) and (\ref{eq:jump_s}), to occur in the time 
interval $(t, t + d t ]$ is given by $(i =1,2)$
\begin{equation}
\hspace*{-0.15 cm} p_{\rm i}(t) \!=\! 
\langle \hat J_i^\dagger \hat J_i  \rangle_t \, d t = 
\kappa_i \, {\rm Tr} \left[ \hat \rho (t)  \hat a^\dagger \hat a \right]  d t = \kappa_i |\beta(t)|^2 
d t
\label{eq:p_jump_i}
\end{equation}
and
\begin{equation}
p_{\rm s}(t) \!=\! 
\langle \hat J_{\rm s}^\dagger \hat J_{\rm s}  \rangle_t \, d t = 
\Gamma \, {\rm Tr} \left[ \hat \rho (t)  \hat A_{22} \right] \! d t \!=\! \Gamma |\alpha(t)|^2 
d t \, .
\label{eq:p_jump_s}
\end{equation}
Of course, the increment in the time interval
$dt$ for $p_{\rm yes}(t)$ is equal to
$p_1(t) + p_2(t) + p_{\rm s}(t)$, so that we can write, using Eqs.~(\ref{eq:p_jump_i}), (\ref{eq:p_jump_s}),
and (\ref{eq:gamma_def}),
\begin{equation}
d|\gamma(t)|^2 = d p_{\rm yes}(t) = 
\kappa |\beta(t)|^2 dt + \Gamma |\alpha(t)|^2 d t\, ,
\label{eq:gamma_diff}
\end{equation}
that is again Eq.~(\ref{eq:gamma_deriv}).
The physical meaning of this relation is quite clear. 
When the system is in $|b\rangle$, i.e. with probability $|\beta(t)|^2$, we can have an emission
of a photon from the cavity or an absorption or scattering
by the cavity mirrors (controlled by the parameter $\kappa$). 
When the system is in $|a\rangle$, i.e. with probability $|\alpha(t)|^2$, we can have a photon
spontaneously emitted by the atom (controlled by the
parameter $\Gamma$). 
The related jumps operators
project the system into $|c\rangle$, hence
producing an increment of $|\gamma(t)|^2$. 
Moreover, by integrating 
equation~(\ref{eq:gamma_diff}) 
one gets
\begin{equation}
p_{\rm yes} (t) = |\gamma(t)|^2 =
p_{\rm ext}(t) + p_{\rm abs}(t) + p_{\rm spo}(t) \, ,
\label{eq:gamma_integral}
\end{equation}
where we have defined
\begin{eqnarray}
p_{\rm ext}(t) = \kappa_1 \int_0^t dt'  |\beta(t')|^2 \, ,
\label{eq:p_ext_t} \\
p_{\rm abs}(t) = \kappa_2 \int_0^t dt'  |\beta(t')|^2 \, ,
\label{eq:p_abs_t}
\end{eqnarray}
and
\begin{equation}
\hspace*{-0.3 cm} p_{\rm spo}(t) = \Gamma \int_0^t dt' |\alpha(t')|^2 \, .
\label{eq:p_spon_t}
\end{equation}
The function $p_{\rm ext}(t)$ represents the probability that
a photon is extracted from the cavity in
the time interval $[0, t]$, and $p_{\rm abs}(t)$
the probability that a photon
is absorbed or scattered by the mirrors 
in the same time interval.
Finally, $p_{\rm spo}(t)$ represents
the probability that a spontaneous emission has
occurred in time interval $[0, t]$.

Note that from Eqs.~(\ref{eq:p_ext_t}), 
(\ref{eq:p_abs_t}), and (\ref{eq:p_spon_t}) 
it follows that $p_{\rm ext}(t)$, $p_{\rm abs}(t)$,
and $p_{\rm spo}(t)$ have to be monotonically increasing
functions: the longer one waits, the larger 
is the probability that a photon has leaked out of
the cavity, or is absorbed or scattered by the mirrors, 
or a spontaneous emission has occurred.
Moreover, if we wait long enough a photon is certain
to be emitted in one of the three ways,
so that $\lim_{t \to \infty} |\gamma(t)|^2 =1$.
In this case $p_{\rm ext}(t)$ does not reach asymptotically the value $1$,
due to the presence of spontaneous emissions and
mirror absorption or scattering.
If we take the limit $t \to \infty$
of Eq.~(\ref{eq:gamma_integral}), we get
\begin{equation}
p_{\rm ext}(\infty) + p_{\rm abs}(\infty) + p_{\rm spo}(\infty) 
= 1 \, .
\label{eq:p_infty}
\end{equation}

In order to determine $\alpha(t)$ and
$\beta(t)$ we have to solve the nonunitary Schr\"odinger equation, cf. Eqs.~(\ref{eq:schr_non_unitary})
and~(\ref{eq:n_H_Hamiltonian}). 
This brings us to consider the following linear system of
differential equations,
\begin{equation}
\left\{ 
\begin{array}{ll}
\dot \alpha(t) 
= -i \left( \Delta - i \frac{\Gamma}{2} \right)\alpha (t) - i g \beta(t) \, , \\
\dot \beta(t) = - i g \alpha(t) - \frac{\kappa}{2} \beta(t) \, .
\end{array}
\right.
\label{eq:sys_diff_eq_1}
\end{equation}
For the initial conditions $\alpha(0) \!=\!1$ and 
$\beta(0) \!=\! 0$, and defining
\begin{equation}
\Omega \equiv \sqrt{\frac{\kappa^2}{4} - 4g^2 - i \kappa \left(\Delta -i \frac{\Gamma}{2}\right) - \left(\Delta - i \frac{\Gamma}{2}\right)^2 } \, , 
\end{equation}
we can write the solutions as
\begin{eqnarray}
&& \hspace*{-0.5cm} {\alpha} (t) \!=\!  \left[\frac{
\kappa/2 - i (\Delta - i \Gamma/2)}{\Omega} \sinh \left(\frac{\Omega t}{2}\right)  \right.
\nonumber \\
&& ~~~~~~~~ + \left. \cosh \left( \frac{\Omega t}{2}\right) \right] \! e^{-[(\kappa+\Gamma)/4 + i \Delta/2]t} , \nonumber \\
&& \hspace*{-0.5cm} {\beta} (t) \!=\! - \frac{2ig}{\Omega} \sinh \left(\frac{\Omega t}{2}\right) e^{-[(\kappa + \Gamma)/4 + i\Delta/2]t}\, .
\label{eq:diff_eq_sol_general_s}
\end{eqnarray}

\begin{figure}
\hspace*{-0.2cm}
   \includegraphics[width=8.5cm]{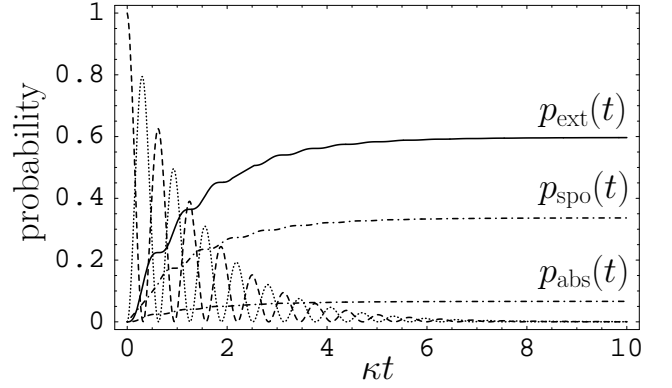} 
\caption{The probabilities $|\alpha(t)|^2$ (dashed line), $|\beta(t)|^2$ (dotted line),
$p_{\rm ext}(t)$ (full line), and $p_{\rm abs}(t)$, $p_{\rm spo}(t)$ (dot-dashed lines) are shown for $2g/\kappa = 10$,
$\Delta/\kappa = 0.1$, $\kappa_1/\kappa = 0.9$,
and $\Gamma/\kappa = 0.5$.}
\label{fig:figure_pra_2}
\end{figure}
Using the solutions given in
Eq.~(\ref{eq:diff_eq_sol_general_s}), one can plot
the probabilities to find at time $t$ the system in 
$|a\rangle$ or in $|b\rangle$,
i.e. $|\alpha(t)|^2$ and $|\beta(t)|^2$, respectively, as well as $p_{\rm ext}(t)$, $p_{\rm abs}(t)$
and $p_{\rm spo}(t)$, as given by Eqs.~(\ref{eq:p_ext_t})--(\ref{eq:p_spon_t}).
In Fig.~\ref{fig:figure_pra_2} we show these functions
for the parameters $2g/\kappa = 10$,
$\Delta/\kappa = 0.1$, $\kappa_1/\kappa =0.9$,
and for the realistic choice of $\Gamma / \kappa = 0.5$, cf. Ref.~\cite{Miller:S551}. 
Note that for $t \to \infty$,
$|\alpha(t)|^2 \to 0$,  $|\beta(t)|^2 \to 0$,
and $p_{\rm ext}(t) + p_{\rm abs}(t) + p_{\rm spo}(t)=|\gamma(t)|^2 \to 1$, as it is expected from 
Eqs.~(\ref{eq:diff_eq_sol_general_s}) and 
(\ref{eq:gamma_def}).

Let us consider $\Delta \!=\!0$ and $g \gg \kappa, \Gamma$. 
From Eq.~(\ref{eq:diff_eq_sol_general_s}) one immediately obtains 
\begin{eqnarray}
&& |\alpha(t)|^2 = \cos^2(gt) e^{-(\kappa+\Gamma) t /2} \, , \nonumber \\
&& |\beta(t)|^2 = \sin^2(gt) e^{-(\kappa+\Gamma) t /2}\, .
\label{eq:diff_eq_sol_a_2}
\end{eqnarray}
In this case the system undergoes damped Rabi oscillations between $|a\rangle$ and $|b\rangle$ with frequency $2g$.
From Eq.~(\ref{eq:diff_eq_sol_a_2}) one easily obtains,
using Eqs.~(\ref{eq:p_ext_t})--(\ref{eq:p_spon_t}),
\begin{eqnarray}
&& \hspace*{-0.8cm} p_{\rm ext} (t) \!=\!
\kappa_1 \!\int_0^t dt'  |\beta(t')|^2
\!=\! \frac{\kappa_1}{\kappa + \Gamma} \left( 1 \!-\! e^{-(\kappa+\Gamma)t/2} \right) , \\
&& \hspace*{-0.8cm} p_{\rm abs} (t) \!=\!
\kappa_2 \! \int_0^t dt'  |\beta(t')|^2
\!=\! \frac{\kappa_2}{\kappa + \Gamma} \left( 1 \!-\! e^{-(\kappa+\Gamma)t/2} \right) ,
\end{eqnarray}
and 
\begin{equation}
\hspace*{-0.1cm}
p_{\rm spo} (t) \!=\!
\Gamma \! \int_0^t dt'  |\alpha(t')|^2
\!=\! \frac{\Gamma}{\kappa + \Gamma} \left( 1 \!-\! e^{-(\kappa+\Gamma)t/2} \right) \, .
\end{equation}
From Eq.~(\ref{eq:gamma_integral})
we then get
\begin{equation}
p_{\rm yes}(t) = |\gamma(t)|^2 = 
1 - e^{-(\kappa + \Gamma)t/2} \, ,
\label{eq:p_yes_2}
\end{equation}
that shows a simple exponential behavior.
For $t \!\to\! \infty$
we have 
$p_{\rm ext} (t) \!\to\! \kappa_1 / (\kappa + \Gamma)$,
$p_{\rm abs} (t) \!\to\! \kappa_2 / (\kappa + \Gamma)$,
$p_{\rm spo} (t) \!\to\! \Gamma / (\kappa +\Gamma)$,
and $ |\gamma(t)|^2 \!\to\! 1$.

\section{Single-photon wave packet}
\label{section:2}

The analysis performed here, using a quantum trajectory
approach, is implicitly based on an unraveling of the master equation~(\ref{eq:master_1}) for the case of direct photoelectric detection of the field emitted from the cavity~\cite{Carmichael}. 
In experiments one uses a large number of photodetections to
recover the properties of the electromagnetic field,
also in the case of single-photon sources,
see, e.\,g., Refs.~\cite{Kuhn:067901, Keller:1075}.
Properties like wave packet duration or wave packet bandwidth 
are analyzed
with an ensemble of photons and cannot be determined from a measurement on just a single photon.
In this respect it is important to carefully  
describe the arrival of photons at the photodetector. 

For this purpose it is convenient,
following the approach 
of~\cite{Blow:4102, Legero:797, Legero:253}, to choose spatiotemporal modes for characterizing the single-photon wave
packet.
A stream of single photons emitted one after the other can be described by a state vector $|1_{\xi_i}\rangle$, where 
\begin{equation}
|1_{\xi_i}\rangle = \hat c_{\xi_i}^\dagger |0\rangle \, .
\label{eq:1_xi_i}
\end{equation}
Here $\hat c_{\xi_i}^\dagger$ is the creation operator for photons of spatiotemporal mode $\xi_i(t)$
defined as
\begin{equation}
\hat c_{\xi_i}^\dagger = \int_{0}^{\infty} dt \, \xi_i(t) 
\hat b^\dagger (t) \, ,
\label{eq:c_def}
\end{equation}
where $\xi_i(t)=0$ for $t<0$, and
$\hat b^\dagger(t)$ is given by
\begin{equation}
\hat b^\dagger(t) = \frac{1}{\sqrt{2 \pi}} \int d\omega \, b^\dagger(\omega) e^{i \omega t} \, ,
\label{eq:b_daga_t}
\end{equation}
that is, the Fourier transform of the operator $\hat b^\dagger(\omega)$, the
creation operator of quanta of a monochromatic wave of frequency $\omega$
in free space. From the relation
\begin{equation}
\left[ \hat b (\omega), \hat b^\dagger(\omega^{'}) \right] = 
\delta(\omega - \omega^{'}) \, ,
\end{equation}
and from Eq.~(\ref{eq:b_daga_t}), it follows that
\begin{equation}
\left[ \hat b (t), \hat b^\dagger(t') \right] = 
\delta(t - t') \, .
\label{eq:comm_t}
\end{equation}
Given that the photon is in the mode $\xi_i$, i.e. it is described by the normalized function $\xi_i(t)$, 
according to
\begin{equation}
\int_{0}^{\infty} dt \, |\xi_i (t)|^2 = 1 \, ,
\label{eq:norm_xi}
\end{equation} 
it is possible to construct a complete orthonormal set $\{ \xi_j(t)\}$ of functions 
where
\begin{equation}
\int dt \, \xi_i(t)\xi_j^{*}(t) = \delta_{ij} \, ,
\label{eq:orthon}
\end{equation}
and
\begin{equation}
\sum_i \xi_i^{*}(t) \xi_i(t') = \delta(t - t') \, .
\end{equation}
From Eqs.~(\ref{eq:comm_t}) and (\ref{eq:orthon}) it is immediate to show that
\begin{equation}
\left[ \hat c_{\xi_i}, \hat c_{\xi_j}^\dagger \right] = \delta_{ij} \, ,
\end{equation}
so that the operators defined by Eq.~(\ref{eq:c_def}) using the complete 
set of orthonormal functions $\{ \xi_j(t)\}$ represent a set of independent bosons, and $\hat c_{\xi_i}$ can be used to construct number states in the usual way,
\begin{equation}
|n_{\xi_i} \rangle = \frac{1}{\sqrt{n !}} \left[ \hat c_{\xi_i}^\dagger \right]^n |0\rangle \, .
\end{equation}

The spatiotemporal mode function ${\xi_i}(t)$ is composed of an amplitude
envelope $\epsilon_i(t)$ and a phase $\phi_i(t)$,
\begin{equation}
\xi_i (t) = \epsilon_i(t) e^{i \phi_i(t)} \, .
\label{eq:xi}
\end{equation}
According to Eq.~(\ref{eq:norm_xi}), the normalization 
reads as
\begin{equation}
\int_{0}^{\infty} dt \, \epsilon^2_i(t) = 1 \, .
\label{eq:norm_epsilon}
\end{equation}
If we now define the flux operator in units of photons per 
unit time,
\begin{equation}
\hat f (t) = \hat b^\dagger (t) \hat b(t) \, ,
\label{eq:flux_op}
\end{equation}
using the inverse relation of Eq.~(\ref{eq:c_def}),
$\hat b (t) = \sum_i \xi_i (t) \hat c_{\xi_i}$, 
we can write Eq.~(\ref{eq:flux_op}) as
\begin{equation}
\hat f (t) = \sum_i \sum_j \xi_i^{*} (t) \xi_j(t) \hat c_{\xi_i}^\dagger
\hat c_{\xi_j} \, .
\label{eq:flux_op_1}
\end{equation}
When no extra losses, such as
spontaneous emissions out the side of
the cavity or mirrors' absorption or scattering,
are considered, the density 
operator of the cavity output field for a photon
in the mode $\xi_i$ is, in the Heisenberg picture,
$\hat \rho_{\rm out} = |1_{\xi_i} \rangle \langle 1_{\xi_i}|$,
cf. Ref.~\cite{Legero:253}.
When extra losses are included, 
the density operator of the cavity output field
is given by the statistical mixture
\begin{equation}
\hat \rho_{\rm out} = 
p_{\rm ext}(\infty)
|1_{\xi_i} \rangle \langle 1_{\xi_i}|
+ \left[ 1 - p_{\rm ext}(\infty) \right] |0\rangle \langle 0| \, .
\label{eq:rho_out_s}
\end{equation}
Note that $1 - p_{\rm ext}(\infty) = p_{\rm abs}(\infty)
+ p_{\rm spo}(\infty)$,
consistently with the fact that the zero-field contribution
is related to the spontaneous emissions out the side of
the cavity or to mirrors' absorption or scattering.
The probability density distribution
of measuring the photon at a given time $t$ is then
\begin{eqnarray}
P_{\xi_i}(t) = {\rm Tr} \left[ \hat \rho_{\rm out} \hat f(t)  \right] 
= p_{\rm ext}(\infty) \epsilon_i^2 (t) \, .
\label{eq:P_i_t} 
\end{eqnarray}
Of course, integrating Eq.~(\ref{eq:P_i_t}),
and using Eq.~(\ref{eq:norm_epsilon}) we have
\begin{equation}
P_{\rm tot} = \int_{0}^{\infty} \! dt \, P_{\xi_i}(t) = 
p_{\rm ext}(\infty) 
\int_{0}^{\infty} \! dt \, \epsilon_i^2 (t) = p_{\rm ext}(\infty) \, ,
\label{eq:P_i_tot} 
\end{equation}
as it is expected.

Let us consider a photon in the mode $\xi_i$,
whose amplitude envelope $\epsilon_i(t)$
does not change significantly in
the detection time resolution $T$.
The response probability of the detector within a time interval $[t-T/2, t+T/2]$ is then, using
Eq.~(\ref{eq:P_i_t}), given by
\begin{equation}
P_{\rm D}(t) \!=\! \int_{t - T/2}^{t + T/2} 
\!\!\!dt' P_{\xi i}(t') 
\simeq
p_{\rm ext}(\infty) \epsilon_i^2(t) \, T  .
\label{eq:P_i_D}
\end{equation}
In the case of a detector of quantum efficiency $\eta$,
Eq.~(\ref{eq:P_i_D}) becomes
\begin{equation}
P_{\rm D}(t)= \eta \, p_{\rm ext}(\infty) 
\epsilon_i^2(t) \, T \, .
\label{eq:P_i_D_eta}
\end{equation}
In a usual experiment a large number of photodetection events are accumulated to obtain $P_{\rm D} (t)$, and from these measurements one gets $\epsilon_i(t)$.
This consideration 
is important because it tells us how one can obtain 
the amplitude envelope of the mode function
within a quantum trajectory 
formalism. 
The probability to measure between time
$t - T/2$ and $t + T/2$ a ``click" at the detector 
is equal to the probability to have a jump $\hat J_1$ in the
same time interval, so that using 
Eq.~(\ref{eq:p_jump_i}), we get, in the case of a detector
of efficiency
$\eta$,
\begin{equation}
P_{\rm D} (t) = \eta \kappa_1  {\rm Tr} \left[ \hat \rho (t)  \hat a^\dagger \hat a \right] T = 
\eta \kappa_1 |\beta(t)|^2 T \, .
\label{eq:D_jumps}
\end{equation}
Comparing this with
Eq.~(\ref{eq:P_i_D_eta}) we obtain
\begin{equation}
\epsilon_i(t) = \sqrt{\frac{\kappa_1}{p_{\rm ext}(\infty)}} |\beta(t)| \, .
\label{eq:epsilon_i_1}
\end{equation}
Let us check if Eq.~(\ref{eq:norm_epsilon}) is still
fulfilled. 
Using Eq.~(\ref{eq:p_ext_t}) we obtain
\begin{equation}
\int_{0}^{\infty} \! dt \, \epsilon_i^2(t) = \frac{1}{p_{\rm ext}(\infty)}
\lim_{t \rightarrow \infty} \kappa_1 \! \int_0^t \! dt' |\beta(t')|^2
= 1 \, .
\end{equation}

\begin{figure}
\hspace*{-0.2cm}
\includegraphics[width=8.5cm]{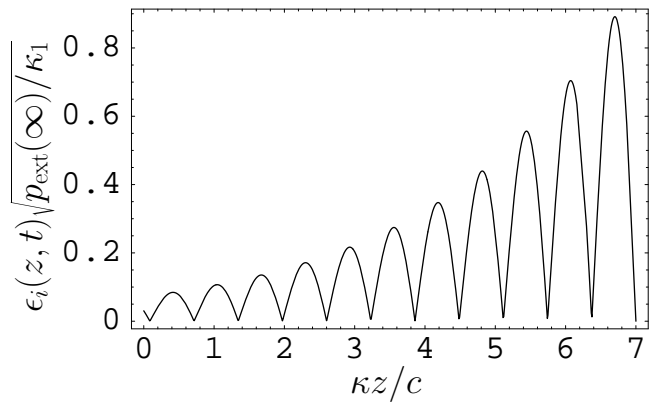} 
\caption{Plot of the amplitude envelope $\epsilon_i(z,t)\sqrt{p_{\rm ext}(\infty)/\kappa_1}$
for the spatiotemporal mode of the cavity output field,
with parameters $2g/\kappa = 10$,
$\Delta/\kappa = 0.1$, $\kappa_1/\kappa = 0.9$,
$\Gamma/\kappa = 0.5$,
and $\kappa t = 7$.}
\label{fig:figure_pra_3}
\end{figure}
To obtain Eq.~(\ref{eq:epsilon_i_1}) we have assumed that the photodetector was positioned just outside the cavity, 
at $z = 0$, so that this equation
gives the amplitude envelope 
of the spatiotemporal mode function at $z = 0$.
Of course we could imagine to position the
detector in an arbitrary position $z$ (with $z>0$).
In this case, because outside the cavity the field
is propagating at the speed of light $c$, the
amplitude envelope $\epsilon_i(z,t)$
is related to the one at position $z = 0$ via the
retarded time $t - z/c$. More precisely,
\begin{equation}
 \epsilon_i(z,t) \!=\! \left\{ 
\begin{array}{ll}
\! \epsilon_i(0,t\!-\!\frac{z}{c}) \!=\! 
\sqrt{\frac{\kappa_1}{p_{\rm ext}(\infty)}}\,
\left|\beta(t\!-\!\frac{z}{c})\right| & \mbox{~$t > \frac{z}{c} > 0$} \\
 \\
\! 0  & \mbox{~~~$t < \frac{z}{c} \, .$}
\end{array}
\right.
\label{eq:epsilon_i_2}
\end{equation}
Note that $|\beta(t)|^2$ represents the probability
to find at time $t$ a photon inside the cavity.
In this respect, Eq.~(\ref{eq:epsilon_i_2}) shows
that the intracavity field dynamics 
determines the structure of the spatiotemporal mode
of the output field.
Moreover, in order to clarify the connection 
between $\epsilon_i(z,t)$ and the probability to emit a
photon, it is useful to consider the following 
equation,
obtained from 
Eqs.~(\ref{eq:epsilon_i_2}) and (\ref{eq:p_ext_t}):
\begin{eqnarray}
&&\int_{z/c}^t dt' \, \epsilon_i^2(z,t') = 
\frac{p_{\rm ext} (t -z/c)}{p_{\rm ext}(\infty)}  ~~~  (t > z/c) \, , 
\end{eqnarray}
where $p_{\rm ext}(t)$ represents the probability
that a photon has leaked out of the cavity in
the time interval $[0, t]$.
Using Eq.~(\ref{eq:epsilon_i_2}) and
the solution given by Eq.~(\ref{eq:diff_eq_sol_general_s}),
in Fig.~\ref{fig:figure_pra_3} we plot the amplitude
envelope $\epsilon_i(z,t)\sqrt{p_{\rm ext}(\infty)/\kappa_1}$ for
the spatiotemporal mode of the cavity
output field for 
the case where the parameters $g$ and $\kappa$ 
are chosen as $2g/\kappa = 10$,
$\Delta/\kappa = 0.1$, $\Gamma/\kappa = 0.5$,
and $\kappa t = 7$.
It is clearly seen that the intracavity dynamics strongly 
modulates the mode structure of the photon wave packet
propagating outside the cavity.

\section{Time control of the photon wave packet}
\label{section:3}

Let us now consider
the case where the interaction of the atom with the cavity-assisted field
has a limited duration, so that it effectively terminates 
at time $\tau$.
To analyze this situation we have obviously
to split the dynamical evolution of the system
in two distinct time intervals, one interval from the initial
time $t_0 =0$ to the time $\tau$, and the second interval 
for times $t$ greater than $\tau$.
For times $t$ such that $ 0 \leq t \leq \tau$, the
evolution is still described by the one previously
analyzed in Sec.~\ref{section:1}, with solutions given by Eq.~(\ref{eq:diff_eq_sol_general_s}). 
For times $t$ with $ t > \tau$, when
the interaction of the atom with the
cavity is set to zero,
the  Hilbert space of the system (cavity field)
reduces to that spanned by the two 
Fock-state vectors $|1\rangle$ and $|0\rangle$
of the cavity field.
At time $\tau$ the cavity field,
obtained by tracing over the atomic states in Eq.~(\ref{eq:rho_t_1}),
is described by
the following density operator:
\begin{equation}
\hat \rho (\tau) = \left[ 1 -  |\beta(\tau)|^2 \right] |0\rangle \langle 0| + |\beta(\tau)|^2 |1 \rangle \langle 1|\, ,
\label{eq:rho_tau}
\end{equation}
where $\beta(\tau)$ is given by Eq.~(\ref{eq:diff_eq_sol_general_s}).

To analyze the dynamical evolution, for $t > \tau$,
of this initial state, we 
follow the procedure
given by the quantum trajectory theory when 
the initial state is not a pure state, but a
statistical
mixture~\cite{Dalibard:580, Dum:4382, Carmichael}.
For the part related to $\left[1 -  |\beta(\tau)|^2 \right] |0\rangle \langle 0|$, one has to start with the state vector
$|\psi(\tau)\rangle = |0\rangle$. 
Because in this case the cavity is
already empty, the evolution simply leaves the cavity
in its vacuum state also at later times, so that we
simply have
\begin{equation}
\hat \rho_1 (t) = |0\rangle \langle 0| \, .
\label{eq:rho_tau_0}
\end{equation}
For the part related to 
$|\beta(\tau)|^2 |1 \rangle \langle 1|$, one has instead 
to start with the state vector $|\psi(\tau)\rangle =|1 \rangle$.
Before a collapse occurs, 
the evolution of the unnormalized state 
$| \bar{\psi}_{\rm no} (t) \rangle = \bar \beta(t) |1 \rangle$
is described by the nonunitary Schr\"odinger equation
\begin{equation}
i \hbar \frac{d}{dt} 
| \bar{\psi}_{\rm no} (t) \rangle  = - i \hbar \frac{\kappa}{2} \hat a^\dagger \hat a  | \bar{\psi}_{\rm no} (t) \rangle .
\label{eq:schr_non_unitary_2} 
\end{equation}
Its solution is
$\bar \beta(t) = e^{-\kappa (t-\tau)/2}$.
If a jump $\hat J_i$, cf. Eq.~(\ref{eq:jump_i}), has occurred at time
$t_{\rm J}$, $t_{\rm J} \in (\tau, t]$, the
wave vector collapses in the state $|0 \rangle$,
\begin{equation}
\hat J_i \, | \bar{\psi}_{\rm no} (t_{\rm J}) \rangle = \sqrt{\kappa_i} \, \hat a \left[ \,
\bar \beta (t_{\rm J})
| 1\rangle \right] \rightarrow |0 \rangle \, .
\label{eq:jump_op_1}
\end{equation}
This tells us that at time $t >\tau$ the density operator for
the part related to the initial state $|\psi(\tau)\rangle =|1 \rangle$ is, using the method of the delay
function~\cite{Dum:4382}, given by
\begin{equation}
\hat \rho_2 (t) =p_{\rm no}(t) |1\rangle \langle 1|+ \left[ 1 - p_{\rm no}(t) \right]  |0\rangle \langle 0| \, ,
\label{eq:rho_1_t}
\end{equation}
where
\begin{equation}
p_{\rm no}(t) = \parallel | \bar{\psi}_{\rm no} (t) \rangle \! \parallel^2
= |\bar \beta(t)|^2 = e^{-\kappa (t-\tau)} \, .
\label{eq:p_no_1}
\end{equation}
Substituting this in Eq.~({\ref{eq:rho_1_t}) one gets
\begin{equation}
\hat \rho_2 (t) = e^{-\kappa (t-\tau)}
|1\rangle \langle 1| + \left[ 1 - e^{-\kappa (t-\tau)} \right]
|0\rangle \langle 0| \, .
\label{eq:rho_tau_1}
\end{equation}
\begin{figure}
\hspace*{-0.2cm}
\includegraphics[width=8.5cm]{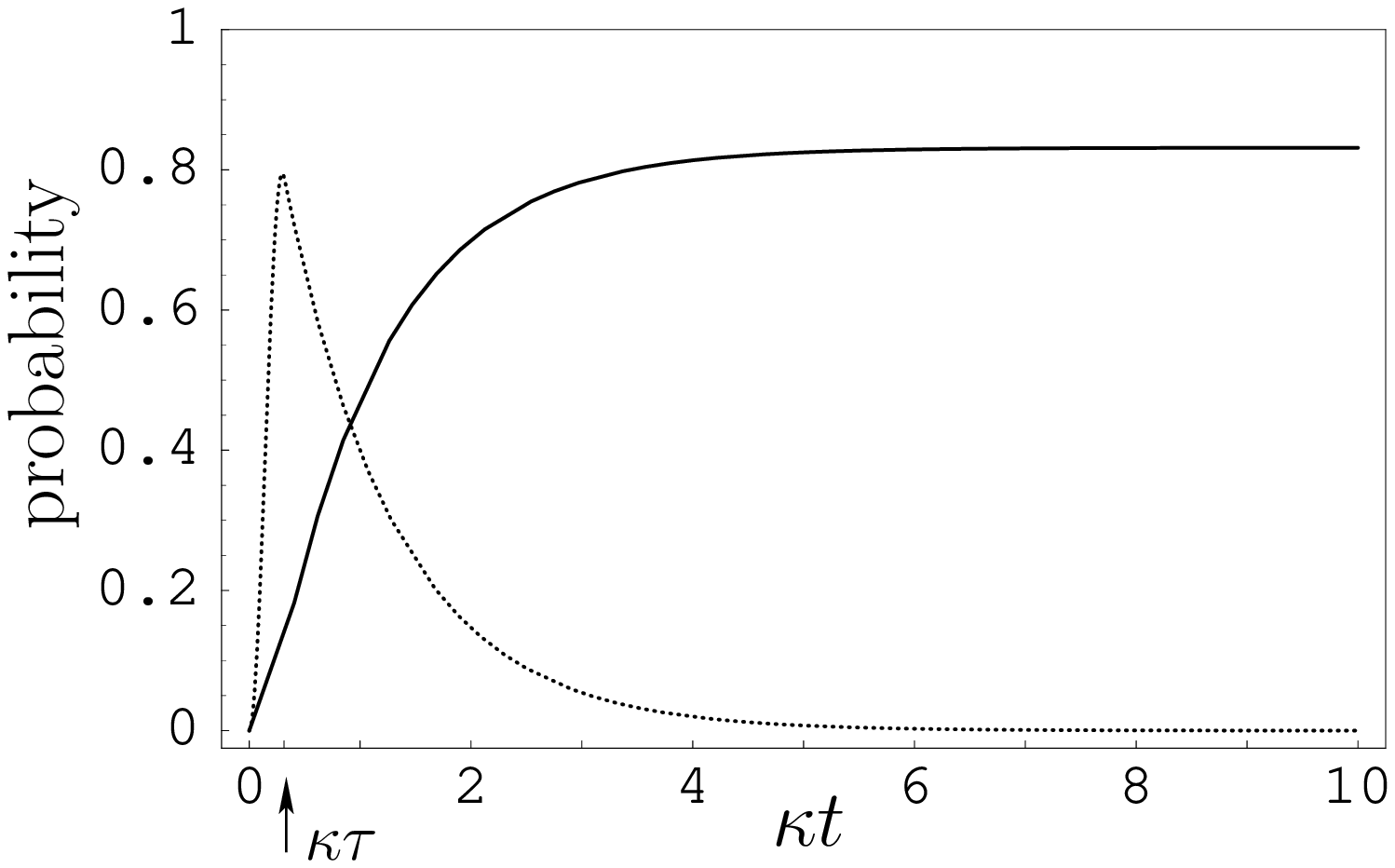} 
\caption{The probabilities $p_{\rm in}(t)$ (dotted line)
and $\bar p_{\rm ext}(t)$ (full line) are shown
for $\tau = \pi/ |\Omega|$,
$2g/\kappa = 10$, $\Delta/\kappa = 0.1$, $\kappa_1/\kappa = 0.9$,
and $\Gamma/\kappa =0.5$.}
\label{fig:figure_pra_4}
\end{figure}

The evolution for the
initial density operator given in Eq.~(\ref{eq:rho_tau})
is, according to the quantum trajectory theory, given by 
\begin{equation}
\hat \rho (t) = \left[ 1 -  |\beta(\tau)|^2 \right] 
\hat \rho_1 (t) + |\beta(\tau)|^2 \hat \rho_2 (t) \, ,
\label{eq:rho_tau_t_1}
\end{equation}
so that we obtain, using Eqs.~(\ref{eq:rho_tau_0}) and (\ref{eq:rho_tau_1}), the result
\begin{eqnarray}
\hat \rho (t) &=& |\beta(\tau)|^2
e^{-\kappa (t-\tau)}
|1\rangle \langle 1| \nonumber \\
&+& \left[ 1 - |\beta(\tau)|^2 e^{-\kappa (t-\tau)} \right]
|0\rangle \langle 0| \, .
\label{eq:rho_tau_2}
\end{eqnarray}
This is the density operator for the intracavity
field for $t > \tau$.
Combining this result with the one for the time interval
$[0,\tau]$, cf. Eq.~(\ref{eq:rho_t_1}),
we can write the probability to find a photon inside
the cavity at an arbitrary time $t$ as
\begin{equation}
p_{\rm in}(t) = \Theta(\tau - t) |\beta(t)|^2 +
\Theta(t - \tau) |\beta(\tau)|^2 e^{- \kappa(t - \tau)} ,
\label{eq:p_in}
\end{equation}
where $|\beta(t)|^2$ and $|\beta(\tau)|^2$
are given by Eq.~(\ref{eq:diff_eq_sol_general_s}),
and $\Theta(t)$ is the unit-step function.

\begin{figure}
\hspace*{-0.2cm}
\includegraphics[width=8.5cm]{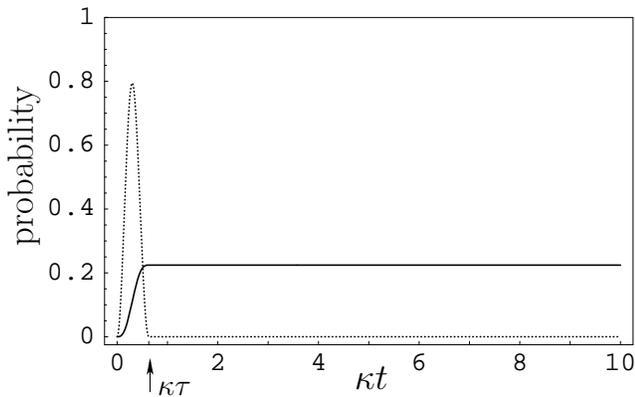} 
\caption{The probabilities $p_{\rm in}(t)$ (dotted line)
and $\bar p_{\rm ext}(t)$ (full line) are shown
for $\tau = 2\pi/ |\Omega|$,
$2g/\kappa = 10$, $\Delta/\kappa = 0.1$, $\kappa_1/\kappa = 0.9$,
and $\Gamma/\kappa =0.5$.}
\label{fig:figure_pra_5}
\end{figure}
Let us now consider the probability $\bar p_{\rm ext}(t)$ 
that a photon is extracted from the cavity in
the time interval $[0, t]$. 
For $t \leq \tau$ this probability is 
equal to $p_{\rm ext}(t)$, cf. Eq.~(\ref{eq:p_ext_t}), with $\beta(t)$ given by Eq.~(\ref{eq:diff_eq_sol_general_s}).
For $t > \tau$ we have a sum of two contributions,
a first one up to time
$\tau$, given by $p_{\rm ext}(\tau)$,
and a second one for the time interval
$[\tau, t]$. This second contribution is
given, using Eqs.~(\ref{eq:p_jump_i}) and
(\ref{eq:rho_tau_2}), by
\begin{eqnarray}
p_{\rm ext}^{\tau}(t) &=& \int_{\tau}^t dt' 
\langle \hat J_1^\dagger \hat J_1  \rangle_{t'} =
 \kappa_1 \!\int_{\tau}^t dt' 
|\beta(\tau)|^2 e^{-\kappa (t'-\tau)} \nonumber \\
&=& \frac{\kappa_1}{\kappa}|\beta(\tau)|^2 \left[ 1 - e^{-\kappa (t-\tau)} \right] \, .
\label{eq:p_ext_1}
\end{eqnarray}
We can now combine these results and write, for an arbitrary 
time $t$,
the probability $\bar p_{\rm ext} (t)$ as  
\begin{equation}
\bar p_{\rm ext} (t) \!=\! 
\Theta(\tau - t) p_{\rm ext}(t) + \Theta(t - \tau)
\left[ p_{\rm ext}(\tau) \!+\! p_{\rm ext}^{\tau}(t) \right] . \label{eq:p_em}
\end{equation}
From this equation one gets that,
for $t \to \infty$, the extraction probability
is equal to $\bar p_{\rm ext}(\infty) = p_{\rm ext}(\tau) + (\kappa_1/\kappa) |\beta(\tau)|^2$.
This relation can be rewritten,
using Eqs.~(\ref{eq:gamma_def}) and (\ref{eq:gamma_integral}),
as
\begin{equation}
\bar p_{\rm ext}(\infty) =
1 - |\alpha(\tau)|^2 - p_{\rm spo}(\tau) 
- p_{\rm abs}(\infty) \, ,
\label{eq:p_ext_infty_tau}
\end{equation}
where $p_{\rm abs}(\infty) = p_{\rm abs}(\tau) + (\kappa_2/\kappa) |\beta(\tau)|^2$, and $p_{\rm abs}(\tau)$,
$p_{\rm spo}(\tau)$ are given by Eq.~(\ref{eq:p_abs_t}) and
Eq.~(\ref{eq:p_spon_t}), respectively.
Eq.~(\ref{eq:p_ext_infty_tau}) shows that the extraction probability is not, in general, asymptotically reaching
the value one.
This reflects the fact that at 
time $\tau$ the atom can be, 
with probability $|\alpha(\tau)|^2$, in its excited state.
If the interaction is set to zero when the atom is in its
excited state, then, obviously, no photon extraction
can anymore occur from the empty cavity.
Moreover, also the contribution due to spontaneous emissions
up to time $\tau$, $p_{\rm spo}(\tau)$, 
and the total absorption probability, $p_{\rm abs}(\infty)$,
remove photons from the extracted output channel.
Note that from Eq.~(\ref{eq:p_ext_infty_tau}) one obtains,
for $\tau \to \infty$, Eq.~(\ref{eq:p_infty}),
as it is expected, being $|\alpha(\tau)|^2 \to 0$.

\begin{figure}
\hspace*{-0.2cm}
\includegraphics[width=8.5cm]{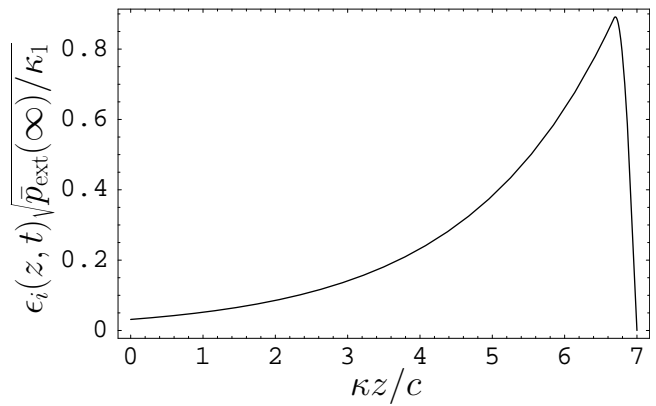} 
\caption{Plot of the amplitude envelope
$\epsilon_i(z,t)\sqrt{\bar p_{\rm ext}(\infty)/\kappa_1}$ 
for the spatiotemporal mode of the
cavity output field, for $\tau = \pi/|\Omega|$,
$\kappa t = 7$, $2g/\kappa = 10$, $\Delta/\kappa = 0.1$,
$\kappa_1/\kappa = 0.9$, and $\Gamma/\kappa =0.5$.}
\label{fig:figure_pra_6}
\end{figure}
Let us now analyze the dynamical evolution
of the system for the following two cases.
In the first case we consider that
the atom leaves the cavity at time
$\tau$ equal to the first half Rabi cycle, 
i.e. at $\tau = \pi/ |\Omega|$, cf. Eq.~(\ref{eq:diff_eq_sol_general_s}).
In the second case we consider that
the atom leaves the cavity at time
$\tau$ equal to the first Rabi cycle, 
i.e. at $\tau = 2\pi/ |\Omega|$. 
For $\tau = \pi/ |\Omega|$,
using Eqs.~(\ref{eq:p_in}) and (\ref{eq:p_em}),
we plot in Fig.~\ref{fig:figure_pra_4} the behavior 
of the probabilities
$p_{\rm in}(t)$ and $\bar p_{\rm ext}(t)$, respectively. 
Note that for the parameters used in this
case we have $|\alpha(\tau)|^2 \simeq 0$.
For $\tau = 2\pi/ |\Omega|$, the behavior of 
the probabilities $p_{\rm in}(t)$ and $\bar p_{\rm ext}(t)$
is plotted in Fig.~\ref{fig:figure_pra_5},
using Eqs.~(\ref{eq:p_in}) and (\ref{eq:p_em}).
Because in this case $|\beta(\tau)|^2 \simeq 0$,
one sees from Eq.~(\ref{eq:p_in}) that the probability
to find a photon inside the cavity at time $t > \tau$
is constant and, approximately, equal to zero.
From Eq.~(\ref{eq:p_em}) the probability 
$\bar p_{\rm ext}(t)$
has, for $t>\tau$, the constant value
$p_{\rm ext}(\tau)$.
Moreover, in this case $|\alpha(\tau)|^2$
is not negligible 
because the interaction was switched off when the atom had
a significant probability to be found in its excited
state. 

We consider now the problem to determine the
single-photon pulse shape in the case of a 
short-term atom-field interaction.
Because for $t > \tau$
the probability to measure between time
$t - T/2$ and $t + T/2$ a ``click" at a photodetector of efficiency
$\eta$, positioned at $z=0$, is given by 
$P_{\rm D}(t) =\eta \kappa_1 |\beta(\tau)|^2 e^{-\kappa(t - \tau)} T$,  
we can write
the amplitude envelope of the spatiotemporal mode
function as
\begin{equation}
\sqrt{\frac{\bar p_{\rm ext}(\infty)}{\kappa_1}}\epsilon_i(t) = 
\Theta(\tau - t) |\beta(t)|
+ \Theta(t - \tau)|\beta(\tau)|
e^{-\frac{\kappa}{2} (t - \tau)} \, ,
\label{eq:epsilon_i_3}
\end{equation}
where $\beta(t)$ and $\beta(\tau)$ are given by Eq.~(\ref{eq:diff_eq_sol_general_s}), 
and $\bar p_{\rm ext}(\infty)$ by Eq.~(\ref{eq:p_ext_infty_tau}).
The factor $\sqrt{\bar p_{\rm ext}(\infty)/\kappa_1}$
is needed in order for the spatiotemporal mode function 
to be properly normalized accordingly to Eq.~(\ref{eq:norm_epsilon}).
If we now define the retarded time $t_{\rm r} \equiv t - z/c$,
we can generalize Eq.~(\ref{eq:epsilon_i_3}) 
as
\begin{eqnarray}
\sqrt{\frac{\bar p_{\rm ext}(\infty)}{\kappa_1}}\epsilon_i(z,t)
&=& 
\Theta(\tau - t_{\rm r}) |\beta(t_{\rm r})| \nonumber \\
&+& \Theta(t_{\rm r} - \tau)
|\beta(\tau)| e^{-\frac{\kappa}{2} (t_{\rm r} - \tau)} \, ,~~~~~
\label{eq:epsilon_i_4}
\end{eqnarray}
for $t > z/c > 0$, and,
obviously, $\epsilon_i(z,t) = 0$ for $ t < z/c$.

\begin{figure}
\hspace*{-0.2cm}
\includegraphics[width=8.5cm]{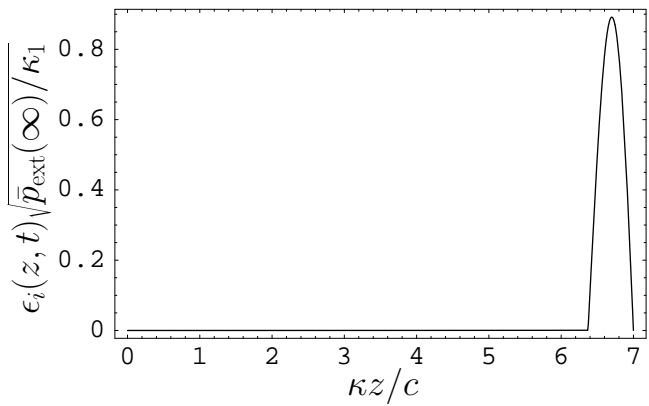} 
\caption{Plot of the amplitude envelope 
$\epsilon_i(z,t)\sqrt{\bar p_{\rm ext}(\infty)/\kappa_1}$
for the spatiotemporal mode of the
cavity output field, for $\tau = 2\pi/|\Omega|$,
$\kappa t = 7$, $2g/\kappa = 10$ and $\Delta/\kappa = 0.1$,
$\kappa_1/\kappa = 0.9$, and $\Gamma/\kappa =0.5$.}
\label{fig:figure_pra_7}
\end{figure}
Using Eq.~(\ref{eq:epsilon_i_4}) one can obtain, for example,
the amplitude envelope of the
spatiotemporal mode for the two cases above considered,
i.e. for $\tau = \pi/|\Omega|$ and for $\tau = 2\pi/|\Omega|$, respectively.
In Fig.~\ref{fig:figure_pra_6} we consider the case where the
interaction time $\tau$ is chosen as $\tau = \pi/|\Omega|$, and 
$\kappa t = 7$.
After $|\beta(t)|^2$, the probability to
find a photon inside the cavity reaches its maximum values at time $\tau$, the interaction with the atom is set to zero,
and one observes an exponential decay regulated by the
photon escape and absorption rate $\kappa$. This behavior is
mapped in the amplitude envelope shape of the 
extracted spatiotemporal
mode, as can be clearly seen in Fig.~\ref{fig:figure_pra_6}.
If we consider now the case where the
interaction time $\tau$ is chosen as $\tau = 2\pi/|\Omega|$,
using Eq.~(\ref{eq:epsilon_i_4}) we obtain,
for $\kappa t = 7$, the function plotted in 
Fig.~\ref{fig:figure_pra_7}.
Here, $|\beta(t)|^2$, after reaching its maximum value, returns practically to zero, for $\tau = 2\pi/|\Omega|$. At this point the interaction with the atom is switched off,
so that the cavity is, practically, left with no photon inside,
so that 
no 
photon can be extracted at later times. 
This dynamics is clearly mapped in the amplitude
envelope shape of the spatiotemporal mode, 
producing a short pulse,
of length $2 \pi c/|\Omega|$, as can be seen in Fig.~\ref{fig:figure_pra_7}.

Finally, we consider 
the amplitude envelope for the spatiotemporal mode of the
cavity output field for an interaction time $\tau$
arbitrary chosen so that $\kappa \tau \!=\! 2.2$. In Fig.~\ref{fig:figure_pra_8},
using Eq.~(\ref{eq:epsilon_i_4}), we
plot the function $\epsilon_i(z,t)
\sqrt{\bar p_{\rm ext}(\infty)/\kappa_1}$
in the region $17 \leq \kappa z/c \leq 20$.
In this plot the amplitude envelope 
shows a behavior that is intermediate
between the one depicted in
Fig.~\ref{fig:figure_pra_6} and the one in Fig.~\ref{fig:figure_pra_3}.
After approximately three and a half Rabi cycles,
the interaction is switched off, and the cavity field
simply decays with an exponential behavior regulated 
by $\kappa$. To clearly see the effects
of spontaneous emissions
we also plot in Fig.~\ref{fig:figure_pra_8}
the behavior of the function $\epsilon_i(z,t)
\sqrt{\bar p_{\rm ext}(\infty)/\kappa_1}$ for
the case $\Gamma =0$. We see that the
presence of spontaneous emissions is
not negligible.
\begin{figure}
\hspace*{-0.2cm}
\includegraphics[width=8.5cm]{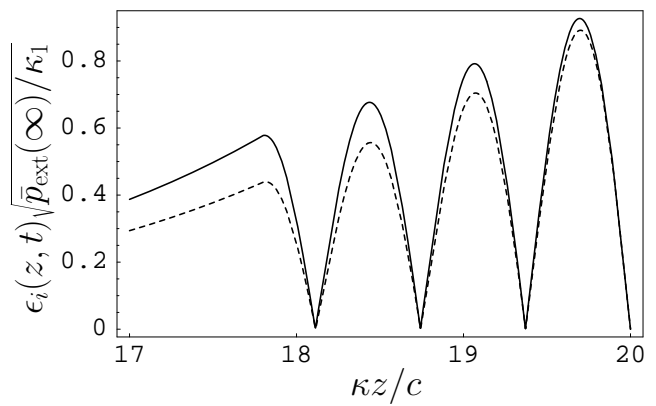} 
\caption{Plot of the amplitude envelope $\epsilon_i(z,t)
\sqrt{\bar p_{\rm ext}(\infty)/\kappa_1}$ 
for the spatiotemporal mode of the
cavity output field in the region $17 \leq \kappa z/c \leq 20$, for $\kappa \tau = 2.2$, $\kappa t = 20$, $2g/\kappa = 10$, $\Delta/\kappa = 0.1$, $\kappa_1/\kappa = 0.9$, and
for $\Gamma = 0$ (full line),
or $\Gamma/ \kappa = 0.5$ (dashed line).}
\label{fig:figure_pra_8}
\end{figure}

In order to realize the time control analyzed here,
let us consider a neutral atom~\cite{Mckeever:1992,Hijlkema:253}
or a trapped ion~\cite{Keller:1075} in an
optical cavity.
With the use of an external laser pulse
it is possible to excite the atom to an 
auxiliary electronic state, to decouple 
the atom from the cavity mode.
In this way the interaction time between the
atom and the cavity can be regulated.
Moreover, for a continuous time-dependent control
of the interaction, a pulsed Raman coupling could be useful,
or the atom could be tuned out of resonance by external 
electric or magnetic fields.
A single-photon wave packet with a defined pulse shape
may be used for a bidirectional atom-photon interface
in a quantum network~\cite{Cirac:3221}. 
This relies on the coherent interaction between the atom and the cavity field, 
provided that the effective coupling exceeds the 
atomic decay rates~\cite{Keller:1607}.

\section{Summary and Conclusions}
\label{conclusions}

The dynamics of an atom-cavity system, which consists of
an initially excited two-level atom
in a lossy cavity, has been analyzed.
The open quantum system under study has been described
by means of a master equation.
By using the quantum trajectory method, we have derived 
analytical solutions for the dynamics of the system.
The effects of unwanted losses, such as absorption and 
scattering by
the cavity mirrors and spontaneous emission
of the atom into field modes out the side of the cavity, have also been considered.
After giving a description of the single-photon wave packet
in terms of spatiotemporal mode functions, we have
connected the probability to measure
a photon in a definite mode structure of the output
field with the photodetection probability. 
In this way the shape of the mode of the extracted
cavity field has been obtained. The mode shape sensitively
depends on the atom-field interaction,
showing a clear 
mapping of the intracavity field dynamics onto the output field. 
The probability
of the mode to carry a one-photon Fock state has been
calculated.
We have also shown that different pulse
shapes of the extracted field can be generated
by controlling the duration of the atom-field
interaction time.

Finally we would like to comment on a fundamental difference of the quantum noise approach under study in comparison with the treatment
of the same problem by macroscopic QED. In the latter method one does not distinguish between the intracavity and the external fields.
There exists only a unique field mode, which covers both the areas inside and outside the cavity. In the 
quantum noise theory on the other hand, the input-output coupling is 
introduced via the interaction of two types of modes, describing the intracavity and the external fields. 
These modes belong to different Hilbert spaces and are therefore commuting. In the description of a unified mode, as in macroscopic QED, there is no hint of the existence of commuting field observables that might describe intracavity and external fields. 
To overcome this basic difference between the two treatments,
it seems reasonable to assume that entanglement between the commuting field modes in the quantum noise theory may replace the noncommutativity in the macroscopic QED in some respect. 
A careful study of this problem requires further investigations.

\section*{ACKNOWLEDGMENTS}
This work was supported by the Deutsche Forschungsgemeinschaft.
The authors thank Thomas Richter for useful discussions.


\newpage

\begin{references}

\bibitem{HarocheRaimond} S.~Haroche and J.-M.~Raimond, {\em Exploring the Quantum} (Oxford University Press, Oxford, 2006).

\bibitem{Walther1} H.~Walther, Fortschr. Phys. {\bf 54},
617 (2006).

\bibitem{Walther2} H.~Walther, B.T.H.~Varcoe, B.G.~Englert,
and T.~Becker, Rep. Prog. Phys. {\bf 69}, 1325 (2006).

\bibitem{Nielsen} M.A.~Nielsen and I.L.~Chuang, {\em Quantum Computation and Quantum Information} (Cambridge University Press, Cambridge, England, 2000).

\bibitem{Haroche:347} S.~Haroche and J.-M.~Raimond,
Adv. At. Mol. Phys. {\bf 20}, 347 (1985).

\bibitem{Gallas:414} J.A.C.~Gallas, G.~Leuchs, H.~Walther, and H.~Figger, Adv. At. Mol. Phys. {\bf 20}, 413 (1985).

\bibitem{Meschede:551} D.~Meschede, H.~Walther, and G.~M{\"u}ller, Phys. Rev. Lett. {\bf 54}, 551 (1985).

\bibitem{Rempe:353} G.~Rempe, H.~Walther, and N.~Klein,
Phys. Rev. Lett. {\bf 58}, 353 (1987).

\bibitem{Brune:1899} M.~Brune, J.M.~Raimond, P.~Goy, L.~Davidovich, and S.~Haroche, Phys. Rev. Lett. {\bf 59}, 1899 (1987).

\bibitem{Thompson:1132} R.J.~Thompson, G.~Rempe, and H.J.~Kimble, Phys. Rev. Lett. {\bf 68}, 1132 (1992).

\bibitem{Monroe:238} C.~Monroe, Nature {\bf 416}, 238 (2002).

\bibitem{Bennett:2724} C.H.~Bennett and P.W.~Shor, IEEE Trans. Inf. Theory {\bf 44}, 2724 (1998).

\bibitem{Luetkenhaus:52304} N.~L{\"u}tkenhaus, Phys. Rev. A {\bf 61}, 52304 (2000).

\bibitem{Cirac:3221} J.I.~Cirac, P.~Zoller, H.J.~Kimble, and
H.~Mabuchi, Phys. Rev. Lett. {\bf 78}, 3221 (1997).

\bibitem{Knill:46} E.~Knill, R.~Laflamme, and G.J.~Milburn, 
Nature {\bf 409}, 46 (2001).

\bibitem{Parkins:3095} A.S.~Parkins, P.~Marte, P.~Zoller, and H.J.~Kimble, Phys. Rev. Lett. {\bf 71}, 3095 (1993).

\bibitem{Hennrich:4872} M.~Hennrich, T.~Legero, A.~Kuhn,
and G.~Rempe, Phys. Rev. Lett. {\bf 85}, 4872 (2000).

\bibitem{Mckeever:1992} J.~McKeever, A.~Boca, A.D.~Boozer, R.~Miller, J.R.~Buck, A.~Kuzmich, and H.J.~Kimble, Science {\bf 303}, 1992 (2004). 

\bibitem{Hijlkema:253} M.~Hijlkema, B.~Weber, H.P.~Specht, 
S.C.~Webster, A.~Kuhn, and G.~Rempe, Nature Physics 
{\bf 3}, 253 (2007).

\bibitem{Wilk:063601} T.~Wilk, S.C.~Webster, H.P.~Specht,
G.~Rempe, and A.~Kuhn, Phys. Rev. Lett. {\bf 98}, 063601 (2007).

\bibitem{Kuhn:067901} A.~Kuhn, M.~Hennrich, and G.~Rempe,
Phys. Rev. Lett. {\bf 89}, 067901 (2002).

\bibitem{Keller:1075} M.~Keller, B.~Lange, K.~Hayasaka, W.~Lange, and H.~Walther, Nature {\bf 431}, 1075 (2004).

\bibitem{Jaynes:89} E.T.~Jaynes and F.W.~Cummings, Proc.~IEEE {\bf 51}, 89 (1963).

\bibitem{Barnett:2444} S.M.~Barnett and P.L.~Knight,
Phys. Rev. A {\bf 33}, 2444 (1986).

\bibitem{Filipowicz:3077} P.~Filipowicz, J.~Javanainen,
and P.~Meystre, Phys. Rev. A {\bf 34}, 3077 (1986).

\bibitem{Puri:3433} R.R.~Puri and G.S.~Agarwal,
Phys. Rev. A {\bf 35}, 3433 (1987).

\bibitem{Kuklinski:3175} J.R.~Kukli\'{n}ski and J.L.~Madajczyk,
Phys. Rev. A {\bf 37}, 3175 (1988).

\bibitem{Risken:346} J.~Eiselt and H.~Risken,
Phys. Rev. A {\bf 43}, 346 (1991).

\bibitem{Cirac:4541} J.I.~Cirac, H.~Ritsch, and P.~Zoller, 
Phys. Rev. A {\bf 44}, 4541 (1991).

\bibitem{Quang:6092} T.~Quang, P.L.~Knight, and V.~Bu\u{z}ek, 
Phys. Rev. A {\bf 44}, 6092 (1991).

\bibitem{Alsing:13} P.~Alsing and H.J.~Carmichael, 
Quantum Opt. {\bf 3}, 13 (1991).

\bibitem{Banacloche:2221} J.~Gea-Banacloche,
Phys. Rev. A {\bf 47}, 2221 (1993).

\bibitem{Shore:1195} B.W.~Shore and P.L.~Knight,
J. Mod. Opt. {\bf 40}, 1195 (1993).

\bibitem{Barnett:2033} S.M.~Barnett and J.~Jeffers,
J. Mod. Opt. {\bf 54}, 2033 (2007).

\bibitem{Knoell:1} L.~Kn\"oll, S.~Scheel and D.-G. Welsch,
{\em Coherence and Statistics of Photons and Atoms}
(Wiley, New York, 2001), chap. 1, quant-ph/0006121.

\bibitem{VogelWelsch} W. Vogel and D.-G. Welsch, {\em Quantum Optics} (Wiley-VCH, Weinheim, 2006), third, revised and extended ed.

\bibitem{Knoell:543} L.~Kn{\"o}ll, W.~Vogel, and D.-G.~Welsch, Phys. Rev. A {\bf 43}, 543 (1991).

\bibitem{Plank:1791} R.W.F.~van der Plank and L.G.~Suttorp, Phys. Rev. A {\bf 53}, 1791 (1996).

\bibitem{Khanbekyan:053813} M.~Khanbekyan, L.~Kn\"oll, D.-G.~Welsch, A.A.~Semenov, and W.~Vogel,
Phys. Rev. A {\bf 72}, 053813 (2005).

\bibitem{Semenov:033803}
A.A.~Semenov, D.Yu.~Vasylyev, W.~Vogel, M.~Khanbekyan, and D.-G.~Welsch, Phys. Rev. A {\bf 74}, 033803 (2006).

\bibitem{Semenov:013807}
A.A.~Semenov, W.~Vogel, M.~Khanbekyan, and D.-G.~Welsch,
Phys. Rev. A {\bf 75}, 013807 (2007).

\bibitem{Khanbekyan:quant-ph} M.~Khanbekyan, 
D.-G.~Welsch, C.~Di Fidio, and W.~Vogel, quant-ph/0709.2998v2. 

\bibitem{Collett:1386} M.J.~Collett and C.W.~Gardiner, Phys. Rev. A {\bf 30}, 1386 (1984).

\bibitem{Gardiner:3761} C.W.~Gardiner and
M.J.~Collett, Phys. Rev. A {\bf 31}, 3761 (1985).

\bibitem{GardinerZoller} G.W.~Gardiner and P.~Zoller, {\em Quantum Noise} (Springer, Berlin, 2004) third ed.

\bibitem{Haake} F.~Haake, {\em Statistical Treatment of Open System by Generalized Master Equations}, (Springer, Berlin, 1973), Vol. 66  in {\em Springer Tracts in Modern Physics}.

\bibitem{Louisell} W.H.~Louisell, {\em Quantum Statistical Properties of Radiation} (Wiley, New York, 1973).

\bibitem{Davies} E.B.~Davies, {\em Quantum Theory of Open Systems} (Academic Press, New York, 1976).

\bibitem{Dalibard:580} J.~Dalibard, Y.~Castin, and K.~M{\o}lmer, Phys. Rev. Lett. {\bf 68}, 580 (1992).

\bibitem{Dum:4382} R.~Dum, A.S.~Parkins, P.~Zoller, and C.W.~Gardiner, Phys. Rev. A {\bf 46}, 4382 (1992).

\bibitem{Carmichael} H.J.~Carmichael, {\em An Open System Approach to Quantum Optics} (Springer, Berlin, 1993)
Vol. m18 of {\em Lecture Notes in Physics, New Series m: Monographs}.

\bibitem{Agarwal:1757} G.S.~Agarwal and R.R.~Puri,
Phys. Rev. A {\bf 33}, 1757 (1986).

\bibitem{Puri:3610} R.R.~Puri and G.S.~Agarwal,
Phys. Rev. A {\bf 33}, 3610 (1986).

\bibitem{Briegel:3311} H.-J.~Briegel and B.-G.~Englert, 
Phys. Rev. A {\bf 47}, 3311 (1993).

\bibitem{Miller:S551} R.~Miller, T.E.~Northup, K.M.~Birnbaum, 
A.~Boca, A.D.~Boozer, and H.J.~Kimble, 
J. Phys. B: At. Mol. Opt. Phys. {\bf 38}, S551 (2005).

\bibitem{Blow:4102} K.J.~Blow, R.~Loudon, S.J.D.~Phoenix, and T.J.~Shepherd, Phys. Rev. A {\bf 42}, 4102 (1990).

\bibitem{Legero:797} T.~Legero, T.~Wilk, A.~Kuhn, and G.~Rempe, Appl. Phys. B {\bf 77}, 797 (2003)

\bibitem{Legero:253} T.~Legero, T.~Wilk, A.~Kuhn, and G.~Rempe, Adv. At. Mol. Opt. Phys. {\bf 53}, 253 (2006).

\bibitem{Keller:1607} M.~Keller, B.~Lange, K.~Hayasaka,
W.~Lange, and H.~Walther, J. Mod. Opt. {\bf 54}, 1607 (2007).

\end{references}
\end{document}